\documentstyle[11pt,aaspp4]{article}

\begin{document}

\title{Efficiency and spectrum of internal $\gamma$-ray burst shocks}
 
\author{D. Guetta\altaffilmark{1,2}, M. Spada\altaffilmark{1,2}, 
E. Waxman\altaffilmark{2}}
\altaffiltext{1}{Osservatorio di Arcetri L. E. Fermi 2 50100 Firenze,
Italy}
\altaffiltext{2}{Department of Condensed Matter Physics, Weizmann
Institute, Rehovot 76100, Israel}

\begin{abstract}

We present an analysis of the Internal Shock Model of GRBs, where
gamma-rays are produced by internal shocks within a relativistic wind.
We show that observed GRB characteristics impose stringent constraints 
on wind and source parameters. We find that a significant fraction, of 
order 20\%, of the wind kinetic energy can be converted to radiation, 
provided the distribution of Lorentz factors within the wind has a 
large variance and provided the minimum Lorentz factor is 
$>\Gamma_{\pm}\approx10^{2.5}L_{52}^{2/9}$, where 
$L=10^{52}L_{52}{\rm erg\,s}^{-1}$ is the wind luminosity.
For a high, $>10\%$, efficiency wind, spectral energy breaks in 
the 0.1 to 1~MeV range are obtained for sources with dynamical time 
$R/c\lesssim1$~ms, suggesting a possible 
explanation for the observed clustering of spectral break energies
in this range. The lower limit $\Gamma_{\pm}$ to wind Lorenz factor
and the upper limit $\approx1(R/10^7{\rm cm})^{-5/6}$~MeV to observed
break energies are set by Thomson optical depth due to $e^{\pm}$ pairs
produced by synchrotron photons.
Natural consequences of the model are
absence of bursts with peak emission energy significantly 
exceeding 1~MeV, and existence of low luminosity
bursts with low, 1~keV to 10~keV, break energies.

\end{abstract}


\section{Introduction}

The widely accepted interpretation of the
phenomenology of $\gamma$-ray bursts (GRBs) is that the
observable effects are due to the dissipation of the kinetic energy
of a relativistically expanding wind, a ``fireball''
(see \cite{Meszaros95} and \cite{Piran96} for reviews).
The discovery
of GRB afterglow emission in X-ray (Costa et al. 1997), 
optical (van Paradijs et al. 1997) and radio wavelengths
(Frail et al. 1997) confirmed 
(Waxman 1997, Wijers, Rees \& M\'esz\'aros 1997), 
standard model predictions of afterglow
(Paczy\'nski \& Rhoads 1993, Kats 1994, M\'esz\'aros \& Rees 1997,
Vietri 1997) that results from the 
collision of the expanding fireball with surrounding medium.

Both the spectrum and temporal dependence of afterglow emission
are consistent with synchrotron emission of electrons accelerated to high
energy at the shock wave driven by the fireball into its surrounding medium.
The rapid temporal variability of $\gamma$-ray emission, i.e. of the GRB
itself, imply, on the other hand, that the GRB is produced by internal 
shocks within the expanding wind (\cite{Woods95,SNP97};
Daigne \& Mochkovitch, 1998;
Panaitescu, Spada \& M\'esz\'aros 1999). Both synchrotron
and inverse-Compton emission from shock accelerated electrons have been
proposed as the GRB emission mechanism. However, synchrotron emission
is favored if the fireball is required to be ``radiatively efficient,'' i.e.
if a significant fraction of the fireball energy 
is required to be converted to $\gamma$-rays (e.g. \cite{Derishev00}).
Observations of GRB970508 afterglow implies a high radiative
efficiency for this GRB
(\cite{FWK00}), and X-ray afterglow
observations imply that high radiative efficiency is required for most GRBs 
(\cite{Freedman00}). 

Observations therefore suggest that the GRB emission is due 
to synchrotron emission of electrons accelerated
by internal fireball shocks. This model faces, however, three
major difficulties. First, it is generally argued 
that if $\gamma$-ray emission
is due to internal shocks within the fireball, then only a small
fraction, $<10^{-2}$, of fireball energy is converted to $\gamma$-rays
(Daigne \& Mochkovitch, 1998; \cite{Kumar,KnP99,Spada}). 
This is in conflict with the high radiative efficiency
implied by afterglow observations. Second, the $\gamma$-ray 
emission of more than 60\% of the GRBs 
observed by BATSE detectors is peaked at photon energies
in the range of
50~keV to 300~keV (Brainerd et al. 1999). It is generally claimed that
clustering of peak emission energy in this narrow energy range requires
fine tuning of fireball model parameters, which should naturally 
produce a much wider range of peak emission energies. 
Finally, there is evidence that, at least in some cases, 
the GRB spectra at low energy is steeper
than expected for synchrotron emission (\cite{Preece98,Frontera00}).
 
The steep low energy spectra has led several authors to consider 
inverse-Compton emission as the source of $\gamma$ radiation
(e.g. Lazzati, et al 2000). However, steep low energy spectra
may also be accounted for within the context of the synchrotron emission
hypothesis by the effects of inverse-Compton suppression at low energies 
(\cite{Derishev00}) and by the contribution to observed $\gamma$-ray
radiation of photospheric fireball emission (e.g. \cite{MnR00}).
As we show here (see discussion in \S4), 
the contribution of latter effect is expected to be significant.

The main goal of the present paper is to determine the constraints imposed
by GRB observations on the relativistic fireball wind parameters, and hence 
on the underlying source producing the wind, under the
assumption that the observed radiation is due to synchrotron emission 
produced in internal wind shocks.
In particular, we address the questions of whether, and under what 
conditions, high radiative efficiency can be obtained, and whether the
clustering of peak emission energies can be naturally explained by the
model.

It has been demonstrated in earlier work
(e.g. \cite{Kobayashi97}; Beloborodov 2000), 
that high radiative efficiency may be obtained
in  the internal shock model provided the variance of the wind Lorentz 
factors distribution is large. Kobayashi et al. (1997) found
$\sim10\%$ efficiency for uniform Lorentz factor distribution with maximum 
to minimum Lorentz factor ratio $\approx10$, and Beloborodov (2000)
demonstrated that the efficiency may approach 100\% for non-uniform 
distributions with wide range of Lorentz factors\footnote{Kobayashi \& Sari 
(2001) similarly find $\sim60\%$ efficiency for non-uniform distribution
of Lorentz factors over the range $10$ to $10^4$.}.
In the analysis presented here we consider several issues which have not 
been addressed in earlier work: 
The constraints imposed on the model by the observed peak emission 
energy distribution, the effect of optical depth due
to $e^\pm$ pairs produced within the fireball wind, and the upper limit  
imposed on the maximum wind Lorentz factor by the acceleration process.
The analysis presented here is also more general, as we study the dependence
of radiative efficiency on a wider set of models parameters. 
We find that the radiative
efficiency is limited to values well below 
the 80\%--90\% derived by
Beloborodov. The inclusion of $e^\pm$ pair optical depth effects
is essential in determining both the efficiency and the peak emission energy.

High radiative efficiency clearly requires that a large fraction of the 
internal energy generated by internal shocks be carried by shock
accelerated electrons, i.e. that the electron and proton energy densities
be close to equipartition. Moreover, high radiative efficiency requires
the magnetic field energy density to be not far below equipartition as well
(e.g. \cite{Derishev00}).
Thus, we assume in most our calculations that 
electrons, protons, and magnetic field energy densities 
are close to equipartition, and focus on the 
study of the dependence of observed $\gamma$-ray
flux and spectrum on wind luminosity and on wind dynamical parameters 
(source size, variability time, Lorenz factor and mass distribution
of shells within the wind). 
For completeness, we demonstarte using our calculations (in \S3.2) that
near equipartition is required in order for model predictions to be
consistent with the observed peak emission energy distribution.

The model outline is presented in \S2. The results of a numerical
analysis of the model are presented and discussed in \S3. 
The main conclusions are summarized in \S4.

\section{Outline of the model}

In the fireball model of GRBs, a compact source, of linear scale
$R_0\sim10^6$~cm, produces a wind characterized by an average luminosity 
$L_w\sim10^{52}{\rm erg\,s}^{-1}$ and mass loss rate $\dot M=L_w/\eta c^2$
($R_0\sim10^6$~cm corresponds to three times the Schwarzschild radius of
a non-rotating, solar mass black hole, and $L_w\sim10^{52}{\rm erg\,s}^{-1}$
is the characteristic luminosity inferred from observations). At small radius, 
the wind bulk Lorentz factor, $\Gamma$, 
grows linearly with radius, until most of the wind energy is converted
to kinetic energy and $\Gamma$ saturates at $\Gamma\sim\eta\sim300$.
$\eta\gtrsim300$ is required to reduce the wind pair-production
optical depth for observed
high energy, $>100$~MeV, photons to less than unity.
If $\eta>\eta_*\approx(\sigma_T L_w/4\pi m_p c^3 R_0)^{1/4}=
2\times10^3(L_{w,52}/R_{0,6})^{1/4}$, where 
$L_w=10^{52}L_{w,52}{\rm erg\,s}^{-1}$ and $R_0=10^6R_{0,6}$~cm,
the wind becomes optically thin at $\Gamma\approx\eta_*<\eta$, and hence
acceleration saturates at $\Gamma\approx\eta_*$ and the remaining wind
internal energy escapes as thermal radiation at $\sim1$~MeV temperature.
Variability of the source on time scale $t_v$, resulting
in fluctuations in the wind bulk Lorentz factor $\Gamma$ on similar
time scale, then leads to internal shocks
in the expanding fireball at a radius
\begin{equation}
R_i\approx\Gamma^2c t_v=3\times10^{11}\Gamma^2_{2} t_{v,-3}
{\rm\ cm},
\label{eq:Ri}
\end{equation}
where $\Gamma=10^{x}\Gamma_{x}$, $t_{v}=10^{-3} t_{v,-3}$~s.
If the Lorentz factor variability within the wind is significant,
internal shocks reconvert a substantial 
part of the kinetic energy to internal energy. It is assumed that
this energy is then radiated as 
$\gamma$-rays by synchrotron (and inverse-Compton) emission of
shock-accelerated electrons.

In this work, we use an approximate model of the
unsteady wind described in the preceding paragraph 
[following \cite{Kobayashi97}; Panaitescu,
Spada, \& M\'esz\'aros 1999; Spada, Panaitescu \& M\'esz\'aros 2000 (SPM00)].
The wind evolution is
followed starting at radii larger than the saturation radius, i.e. 
after the shells have already reached their final 
Lorenz factor following the acceleration phase, and the GRB photon 
flux and spectrum resulting from a series 
of internal shocks that occur within the wind at larger radii
are calculated. 
We consider both synchrotron and the inverse-Compton emission, 
and take into account the effect of $e^\pm$ pair production.

We approximate the wind flow as a set of discrete shells. 
Each shell is characterized by four 
parameters: ejection time $t_j$, where the subscript $j$ denotes the
$j$-th shell, Lorentz factor $\Gamma_j$, mass $M_j$, and width $\Delta_j$. 
Since the wind duration, $t_w\sim10$~s, is much larger than the dynamical
time of the source, $t_d\sim R_0/c$, variability of the wind on a wide range
of time scales, $t_d<t_v<t_w$, is possible. For simplicity, we consider
a case where the wind variability is characterized by a single time scale 
$t_w>t_v>t_d$, in addition to the dynamical time scale of the source $t_d$
and to the wind duration $t_w$.
Thus, we consider shells of initial thickness 
$\Delta_j=c t_d=R_0$, ejected from the source
at an average rate $t_v^{-1}$. The shell 
thickness may increase with time, due to variations in the expansion
Lorenz factor across the shell. The Lorenz factor distribution across
the shell depends on the entropy distribution, which is determined by the
details of the ejection from the source. 
We therefore consider two extreme cases: uniform 
Lorenz factor across the shell, in which case the shell thickness is time 
independent, and order unity variation of LF across the shell, 
$\Delta\Gamma/\Gamma\sim1$, in which case the shell thickness is given 
by $\Delta_j = \max(R_0,R/\Gamma_j^2)$.

We assume
that the Lorenz factor (LF) of a given shell is independent of 
those of preceding shells, and consider 4 different LF distributions:
uniform, modulated, bimodal, and lognormal. 
In the uniform distribution case, the shells' LFs 
are randomly drawn from a uniform distribution over the range  
$\Gamma_{m}$ to $\Gamma_{M}$, where $\Gamma_{m}<\Gamma_{M}$ are time 
independent. In the modulated case, 
$\Gamma_{M}$ varies as $\sin^2(2\pi t_j/t_w)$.
In the bimodal case, the LFs are drawn from a bimodal distribution,
$\Gamma_j=\Gamma_m$ or $\Gamma_j=\Gamma_{M}$ with equal probability. Finally
in the lognormal case, shell LFs are drawn from a
lognormal distribution with an average $\langle\Gamma\rangle=1000$ and 
$\langle\Gamma^2\rangle^{1/2}/\langle\Gamma\rangle=4$, 
truncated at $\Gamma>\eta_*$.

The time intervals $t_{j+1}-t_j$ are drawn randomly from
a uniform distribution with an average value of $t_v$.
The shell masses $M_j$ are chosen to be either independent of $j$, or inversely
proportional to $\Gamma_j$. That is, we consider the two qualitatively 
different possibilities: equal shell mass and equal shell energy. 
The total mass is determined by the constraint 
$\sum_{j=1}^{N} M_j\Gamma_j c^2 =L_w t_w$, where $N=t_w/t_v$.

Once shell parameters are determined, we calculate the radii where  
collisions occur and determine the emission 
from each collision. We assume that following each collision
the two colliding shells merge and continue to expand as a single
shell. 
Under this assumption, the internal energy carried by the merged 
shell following the collision is
\begin{equation}
E_{in}= (M_1\Gamma_1+M_2\Gamma_2-(M_1+M_2)\Gamma)c^2,
\label{eq:Ein}
\end{equation}
where
\begin{equation}
\Gamma=\sqrt{\frac{M_1\Gamma_1+M_2\Gamma_2}
{M_1/\Gamma_1+M_2/\Gamma_2}}
\label{Gmerg}
\end{equation}
is the expansion LF of the merged shell and $M_{1,2}$, 
$\Gamma_{1,2}$ are the masses and the LFs of the colliding shells.
The dynamical efficiency $\epsilon_d$, defined as the fraction
of total kinetic energy converted to internal energy, increases with the
ratio of the fast and slow shell LFs, 
and, for a given ratio, is maximized for equal masses.
The first collisions, for which the differences in the shell
LFs are largest, are the most efficient. 

In each collision a forward (FS) and a reverse (RS) shock are formed,
propagating into forward and backward shells respectively.
The plasma parameters behind each shock are determined by the Taub
adiabat, requiring continuous energy density and velocity across the contact 
discontinuity separating the two shells (Panaitescu \& M\'esz\'aros 1999).
Since the energy density in the downstream regions of
both shocks is similar,  
we divide the internal energy between the reverse and forward shock 
according to the ratio
$E_{f}/E_{r}=\Delta_{f}/k\Delta_{r}$, where $\Delta_{r,f}$ are 
the shell thicknesses prior to the collision and $k$ is a 
factor of the order of few that takes into account the compression 
ratios of the two colliding shells.

The energy released in each shock is distributed among 
electrons, magnetic field and protons
with fractions $\epsilon_e$, $\epsilon_B$ and $1-(\epsilon_e+\epsilon_B)$
respectively. 
We assume that electrons are accelerated by the shocks to a 
a power-law distribution, $dn_e/d\gamma_e\propto\gamma_e^{-p}$ for particle
Lorentz factors $\gamma_{\min}<\gamma_e<\gamma_{\max}$. $\gamma_{\max}$
is the electron Lorenz factor at which the synchrotron cooling time
equals the acceleration time, estimated as the electron Larmor radius divided
by $c$, while $\gamma_{\min}$ is determined by the requirement that the energy
carried by electrons be equal to $\epsilon_e E_{in}$.

The GRB energy spectrum and flux is derived 
by summing the contributions of individual shell collisions.
For each collision, synchrotron and inverse-Compton emission 
by the shock-accelerated electrons is calculated. In order to achieve
high radiative efficiency, electrons must radiatively loose most of their 
energy on a time scale shorter than shock shell crossing time.
Under this condition, the synchrotron spectrum may be approximated as
\begin{equation}
F_{\nu} \propto \left\{ \begin{array}{ll} 
\nu^{-1/2} & \nu<\nu_{sy} \\ \nu^{-p/2} & \nu_{sy}<\nu
\end{array}
\right. \;,
\label{Fnu1}
\end{equation}
where $\nu_{sy}$ is the characteristic synchrotron emission frequency of
electrons with $\gamma\simeq\gamma_{\min}$. 
The inverse Compton spectrum has a similar shape but 
is shifted to higher energy as described in 
SPM00. In determining the inverse-Compton flux, we take
into account the Klein-Nishina suppression of the inverse-Compton 
cross section.

The photons radiated can be scattered by the 
electrons within the shell. The Thomson optical 
depth $\tau_{T}$ for such scattering is evaluated by taking into account 
the electrons that were accelerated but have cooled radiatively while 
the shock crossed the shell, and those within the yet un-shocked part 
of the shell. The optical depth to Thomson scattering may be increased
significantly beyond the value derived taking into account cooled shell
electrons, due to the production of $e^{\pm}$ pairs.
This contribution has not been taken into account in previous
analysis of the internal shock model, since for a uniform distribution
of Lorentz factors, and under the hypothesis of equipartition,
the break energy, $h\nu_{sy}$, in the shell co-moving frame  
is well below the pair production threshold. However,
since the photon spectrum extends to high energy
as a power law with spectral index $-p/2\approx -1$,
there is an equal number of photons per logarithmic
energy intervals, and there may exist a large number of photons
beyond the pair production threshold. The pairs produced by these photons
may contribute significantly to the Thomson optical depth.

In order to take the effect of pair production into account, 
we determine for each collision the photon energy 
$\epsilon_{\rm pairs}$ for which the pair production optical depth
$\tau_{\gamma\gamma}$ equals unity.
All the photons that have energies higher than 
$\max(m_e c^2,\epsilon_{\rm pairs})$ (measured in the shell frame)
are assumed to form pairs, leading to a suppression
of the high energy photon tail. 
We assume that the energy of photons that undergo pair production
is converted to sub-relativistic pairs, which
are taken into account in determining the Thomson optical depth.

We assume that a fraction $(1-{\rm e}^{-\tau_{r,f}})/\tau_{r,f}$, where
$\tau_{r,f}$ are the optical depths of the RS and FS, 
of the photons produced by each shock
escapes and reaches the observer (The RS flux is also absorbed
by the forward shell, and is thus further reduced by a factor 
$e^{-\tau_f}$). The energy of the absorbed photons, as well as the internal
energy which has not been converted to radiation,
are  converted back to ordered 
kinetic energy by adiabatic losses during the shell expansion. During this
process, we take into account the increase in shell thickness by assuming
this thickness grows as the shell's (comoving frame) speed of sound.

Our assumption, that following a collision two shells merger into one shell
which expands with a single (uniform) Lorenz factor, leads to an 
overestimate of the reduction due to collisions of shell Lorenz factor
variance. Clearly, the expansion of the shocked shells due to the
pressure produced by the shocks results in the leading edge of the two
shells moving faster (in the observer frame) than the trailing edge. 
It is therefore clear that the variance in Lorenz factor distribution
following the collision is larger than obtained under our assumption of
a ``complete'' merger. This, in turn, implies that a more detailed calculation
of the shell merger process, which may require dividing the merged shells
into several distinct shells following the collision, will enhance
the overall efficiency of kinetic to thermal energy conversion.
The magnitude of such enhancement
depends on the density and Lorenz factor distributions within the shells prior 
to the collision, and is expected to be of order unity. We therefore do
not consider this effect in the approximate analysis presented in this
paper.

\section{Results and discussion}

In this section we determine the dependence of the wind
radiative efficiency $\epsilon_{\gamma}$, defined as the fraction of
wind energy converted to radiation, and peak emission energy 
$\varepsilon_{p}$, the photon energy at which the maximum
of $\nu f_\nu$ is obtained, on wind model parameters. As explained in
the introduction, we
adopt electron and magnetic field energy fractions
close to equipartition and 
analyze the dependence of $\epsilon_{\gamma}$ and
$\varepsilon_{p}$ on wind luminosity, variability time, source size, 
and on the distribution of wind Lorentz factors and shell masses. 
We use $\epsilon_e$=0.45, $\epsilon_B=0.1$, and 
$p=2$ throughout the paper,
except in the calculations presented in Fig. 4, where smaller
$\epsilon_e$ and $\epsilon_B$ values 
are assumed in order to demonstarte that near equipartitrion
is required by observations. 
We first discuss in \S3.1 the various Lorenz factor distributions 
described in \S2. We then present a detailed discussion of the bimodal
case, which best reproduces observed GRB characteristics, in \S3.2.

\subsection{Comparison of various Lorenz factor distributions}

Fig. 1 shows the results of numerical simulations of the model described
in the previous section. Each panel shows
$\epsilon_{\gamma}$ versus $\varepsilon_p$ for one of the LF 
distributions described in \S2. 
Different points within each panel correspond to
different choices of $\Gamma_m$, $\Gamma_M$ and $t_v$. Results are shown
for $10< \Gamma_m < \Gamma_M < 2500$ and $10^{-4}{\rm\, s}<t_v<1$~s.
We have assumed equal mass shells, 
$R_{0}=10^{6}$~cm, $L_{w}=10^{51}{\rm erg\, s}^{-1}$ and source redshift
$z=1$ for all calculations. 
Results are presented for the two extreme assumptions on shell
width evolution with radius described in \S2, i.e. 
for both constant width $\Delta=R_0$ and
maximal expansion, $\Delta=\max(R_0,R/\Gamma^2)$. 

Based on Fig.1, a uniform LF distribution
can be ruled out since the radiative efficiency is 
small, $\epsilon_{\gamma}< 5$\%. The low radiative efficiency in this
case is due to low dynamical efficiency, i.e. to the fact that only
a small fraction of the kinetic energy is converted to internal
energy in collisions. In order to increase the dynamical efficiency 
it is necessary to either enhance the variance of the
LF distribution of colliding shells at the first set of
collisions (as is the case for bimodal or lognormal distributions) 
or to keep the variance moderate over a wind range of collision radii
(as is the case for the modulated case).

In the modulated case the maximum dynamical efficiency is 40\%.
The first collisions remove the initial random differences and
the merged shells have LFs near the ejection average value
$\langle{\Gamma}\rangle=(\Gamma_{m}+\Gamma_{M})/2$. If the wind is homogeneous
$\langle{\Gamma}\rangle$ is similar for all shells, resulting
in a steady decrease of $\epsilon_d$ during the wind expansion.
If the range of LFs is varying on time scale of order of $t_w$,
$\langle{\Gamma}\rangle$ reflects the initial modulation of $\Gamma_{M}$
and larger radii collisions that are dynamically efficient are still
possible. Thus, a modulation in the wind can solve the
efficiency problem ($\epsilon_{\gamma}\approx 10\%$). However, 
the peak emission energy in this case is  well below the BATSE 
range, since most of the emission originates at large radii, much larger
then the radius of the photosphere $R_\pm$, the radius
at which the wind becomes optically thin to Thomson scattering by
$e^\pm$ pairs [see the discussion preceding Eq. (\ref{eq:pic}) in \S3.2].

The dynamical efficiency in the case of a bimodal LF distribution can reach
50\%, since the Lorentz factors of the colliding shells 
are very different. Contrary to the case of modulated ejection, 
the collisions at the photospheric radius $R_\pm$ dominate 
wind emission, and 
a region of the parameter space exists, where model radiative efficiency and
peak emission
energy are consistent with observations, $\epsilon_\gamma>10\%$ and 
$\varepsilon_p\sim100$~keV. 

The case of a lognormal LF distribution can be considered
as an intermediate case between the random and bimodal ones.
In this case the maximal radiative efficiency expected 
is of the order 10-15\%, much lower than the 80-90\% value
derived in Beloborodov (2000). This is 
due to the pair optical depth, which sets a lower limit of 
$\Gamma_m\gtrsim100$ for an optically thin wind (see \S3.2 below), and
to the constraint imposed 
on the maximum LF by finite source size, $\Gamma_M<\eta_*\sim10^{3.5}$.
These constraints limit the variance of the LF distribution, and hence
the dynamical efficiency.

Considering shells of equal energy, rather than of equal mass, 
the main results do not change. Uniform LF distribution is still  
characterized by low efficiency and modulation leads to low  
peak emission energy. In the case of a bimodal distribution, the efficiency 
is lower for shells of equal energies than for shells of equal mass,
due to an increase in wind optical depth. 
The LF of the merged shell, given by Eq.(\ref{Gmerg}), is reduced
in the equal energy case by a factor $\sqrt{\Gamma_M/\Gamma_m}$ 
with respect to 
the equal mass case. Thus, the average wind LF
is lower, implying smaller collision radii and higher opacity.

>From the analysis of Fig. 1 we conclude that a bimodal distribution of
shell LFs, that is, LF distribution with large variance,
and equal shell mass are favored by observations.
Note, that this conclusion is independent of the assumption of 
near equipartition, using which the above calculations were performed,
since for electron and magnetic field energy fractions below
equipartition the peak emission energy $\epsilon_p$ would be lower 
than obtained above [see also \S3.2, Fig. 4 and Eq. (\ref{eq:emax1}), below].
In what follows we study the bimodal LF case in more detail,
in order to examine the dependence of observable characteristics 
on wind parameters.

\subsection{Bimodal LF distribution}

Figures 2 and 3 present the dependence on $\Gamma_m$ and on $t_{v}$
of peak emission
energy, $\varepsilon_p$, and radiative efficiency, $\epsilon_\gamma$,
respectively, 
for $\Gamma_M=2500$, $L_w=10^{52}{\rm erg\, s}^{-1}$, $R_{0}=10^6$~cm and  
equal shell masses. 
Results are presented in Fig. 2 for the two extreme assumptions on shell
width evolution with radius, $\Delta=R_0$ and
$\Delta=\max(R_0,R/\Gamma^2)$. The efficiency contour plot, shown in Fig. 3,
is similar for both cases.

The dependence of $\varepsilon_p$ on parameters 
can be understood considering the dependence of the dynamical efficiency
and of the optical depth on $\Gamma_m$ and $t_{v}$.
For high values of $\Gamma_m$ ($>300$), both reverse and forward shocks 
are optically thin.  Decreasing $\Gamma_m$ leads to a decrease in the 
initial radius of the collisions, increasing the efficiency up to 15\%
and the peak emission energy up to 0.1--1~MeV. 
At lower values of $\Gamma_{m}$, the efficiency decreases
and steep breaks are observed in the $\varepsilon_p$ contour plot.
These breaks can be understood considering the difference
between the FS and RS efficiencies, due to different
comoving densities of the two colliding shells ($E_{f}/E_{r}\sim
\rho_{c,r}/\rho_{c,f}$). 
In the case of maximal shell expansion, $\Delta=R/\Gamma^2$,
the ratio $E_{f}/E_{r}\sim \Gamma_r/\Gamma_f$, and the FS
dominates the emission. Thus, the break in the $\epsilon_p$ 
contour plot (left panel of Fig. 2) corresponds to the 
region of $\Gamma_m$ and $t_v$ values where the FS, and 
consequently also the RS, become optically thick due to pair
production ($\Gamma_m=50$ to $\Gamma_m=130$ for $t_{v}$ 
varying between 1~s and $10^{-4}$~s respectively).
A small decrease in $\Gamma_m$ results in collisions below 
the photosphere, leading to a steep decrease in peak emission 
energy, to a value corresponding to the second set of collisions, 
which occur at larger radii. These collisions dominate the 
emission and are characterized by a lower efficiency and a lower 
peak emission energy.  
In the case of a constant shell width $\Delta=R_0$, the ratio
$E_{f}/E_{r}\sim \Gamma_f/\Gamma_r$, and the RS dominates the 
emission.  Thus, there are two breaks in the
peak energy contour plot (right panel of fig 2). The first
corresponds to the values of $\Gamma_m$ and $t_v$
where the RS becomes optically thick due to pair production and the
FS starts to dominate the emission. This break represents the
shift of the peak energy from the RS value to the FS one.
The other occurs in the parameter region where the FS becomes optically thick,
and corresponds to the shift of the peak energy from the first to 
the second set of collisions. 

The contour plot of the radiative efficiency in Fig. 3 
shows that values of $\epsilon_{\gamma}$ higher than 10\%
and peak emission energy 
between 0.1 and 1MeV are reached at similar regions in the
$\Gamma_m$-- $t_v$ plane, in accordance with the correlation 
between the peak emission energy and radiative efficiency shown 
in Fig. 1. High efficiency, $>10\%$, is obtained only 
for large $\Gamma_M$, $\Gamma_M\approx\eta_*\approx2\times10^3$, 
i.e. for values close to
the maximum allowed by the finite source size. For lower values 
of $\Gamma_M$, the dynamical efficiency decreases leading to lower 
$\varepsilon_p$ and $\epsilon_{\gamma}$.

In Fig. 4 we present results of calculations similar to those 
presented in Fig. 2, for $\epsilon_e$ and $\epsilon_B$ values below
equipartition. The results demonstarte that the electron and magnetic field
energy fractions can not be far below equipartition,
$\epsilon_e\gtrsim10^{-1}$ and $\epsilon_B\gtrsim10^{-2}$ are required in order
to obtain peak emission energies consistent with observations. These
results are consisitent with our analytic estimates, 
Eqs. (\ref{eq:emax}) and (\ref{eq:emax1}) below.

The contour plot shown in Fig. 2 demonstrates that values of 
$\varepsilon_p$ larger than $\sim1$~MeV can not be obtained. This
result can be understood using the following arguments.
For a collision of shells of thickness $\Delta$ at radius $R$, we have
\begin{equation}
\epsilon_{p}\approx\gamma_{min}^2\frac{h e}{2\pi m_e c}\sqrt{
\frac{2\epsilon_BE_{in}}{R^2\Delta}}.
\label{eq:pic}
\end{equation}
The maximal value of $\varepsilon_p$ is obtained for collisions at the smallest
radius $R$ for which the wind is optically thin. 
The radius of the Thomson photosphere 
due to the electrons present in the original 
fireball is given by
\begin{equation}
R_T\approx\sqrt{\frac{\sigma_T M}{4\pi m_p}}\approx
6\times10^{11}{\rm cm}\, L_{w,52}^{1/2}\,
t_{v,-3}^{1/2}\,\Gamma_{{\rm M},3}^{-1/2},
\end{equation}
where $M$ is the shell mass, and we have approximated
$M=L_wt_{v}/c^2(\Gamma_{\rm m}+\Gamma_{\rm M})$.
$R_{\pm}$, the Thomson photosphere radius 
due to $e^\pm$ pairs resulting from pair production interaction of 
synchrotron photons, can be estimated by
assuming that a significant fraction, $\sim1/2$, of the radiative 
energy is converted to pairs, as typically is the case.
The number density of $e^{\pm}$ is given in this case by
$n_{\pm} \approx \epsilon_{e}E_{\rm in}/(8 \pi m_e c^2 R^{2}\Gamma^2\Delta)$
and 
\begin{eqnarray} 
R_{\pm}\approx &&\sqrt{\frac{3\sigma_T/16}{8\pi m_e c^2}} 
\sqrt{\frac{\epsilon_{e} E_{\rm in}}{\Gamma}}
\nonumber\cr
\approx &&
10^{13}{\rm cm}\,\epsilon_{e}^{1/2}\,
L_{w,52}^{1/2}\,t_{v,-3}^{1/2} \,
(\Gamma_{{\rm M},3}\Gamma_{{\rm m},2})^{-1/4}.
\label{rpmmin}
\end{eqnarray}
Comparing $R_\pm$ and $R_T$, we conclude that the optical depth is
typically dominated by the pairs. 

For fixed $t_v$ and $\Gamma_M$, the  
maximal peak energy is obtained for $\Gamma_m=\Gamma_{m\pm}$ for which the
collision radius $R_i\approx\Gamma_m^2 c t_v$ equals the photospheric radius
$R_\pm$. Solving for $\Gamma_{m\pm}$,
\begin{equation}
\Gamma_{m\pm}\approx\,5\times10^2\,\epsilon_{e}^{2/9}\,
\Gamma_{{\rm M},3}^{-1/9}\,t_{v,-3}^{-2/9}\,L_{w,52}^{2/9},
\label{eq:gpm} 
\end{equation}
and substituting
in Eq. (\ref{eq:pic}) we find
\begin{equation}
\varepsilon_{p}^{\rm max}\approx2 {\rm MeV}\,
\epsilon_{B}^{1/2}\,\epsilon_{e}^{4/3}\,
\Gamma_{{\rm M},3}^{4/3}\,\Delta_6^{-1/2}\,t_{v,-3}^{1/6}\,L_{w,52}^{-1/6},
\label{eq:emax} 
\end{equation}
where we have used
$\gamma_{\rm min}=\sqrt{(\Gamma_M/\Gamma_m)} \epsilon_e m_p/m_e{\rm log}
(\gamma_{\rm max}/\gamma_{\min})$ with 
$\gamma_{\rm max}=100\, \gamma_{\rm min}.$
The dependence of $\varepsilon_p$ on wind luminosity is demonstrated 
in figures 5 and 6. Under the assumption of fixed shell width, $\Delta=R_0$,
$\varepsilon_p^{\max}$ is weakly dependent on $L$, as indicated by
Eq. (\ref{eq:emax}). If the maximum Lorenz factor scales as 
$\Gamma_M\approx\eta_*\propto L^{1/4}$, rather than being independent
of $L$, the slow decrease with
$L$ of $\varepsilon_p^{\max}$ at fixed $t_v$, 
$\varepsilon_p^{\max}\propto L^{-1/6}$, is replaced by a slow increase with
$L$
\begin{equation}
\varepsilon_{p}^{\rm max}\approx4 {\rm MeV}\,
\epsilon_{B}^{1/2}\,\epsilon_{e}^{4/3}\,
\Delta_6^{-5/6}\,t_{v,-3}^{1/6}\,L_{w,52}^{1/6}.
\label{eq:emax1} 
\end{equation}
The dependence on luminosity becomes stronger under the assumption of
maximal expansion. Substituting $\Delta=R/\Gamma^2$ in Eq. (\ref{eq:emax}),
we find $\varepsilon_p^{\max}\propto L^{-5/18}$. The stronger dependence
is also apparent in figures 5 and 6. However, allowing for a scaling
$\Gamma_M\approx\eta_*\propto L^{1/4}$, the peak emission energy 
dependence on $L$ becomes $\varepsilon_p^{\max}\propto L^{7/36}$,
similar to the dependence shown in Eq. (\ref{eq:emax1}).

\section{Conclusions}

We have analyzed a model of GRBs, in which a compact source of linear scale
$R_0$ produces a wind characterized by an average luminosity 
$L_w$ and mass loss rate $\dot M=L_w/\eta c^2$. At small radius, 
the wind bulk Lorentz factor, $\Gamma$, 
grows linearly with radius, until most of the wind energy is converted
to kinetic energy and $\Gamma$ saturates at $\Gamma\sim\eta$.
Variability of the source results
in fluctuations in the wind saturation Lorentz factor $\Gamma$,
leading to internal shocks in the expanding wind. These shocks
reconvert a fraction
of the kinetic energy back to internal energy, which is assumed to be
radiated as 
$\gamma$-rays by synchrotron (and inverse-Compton) emission of
shock-accelerated electrons. 
Since the wind duration, $t_w\sim10$~s, is much larger than the dynamical
time of the source, $t_d\sim R_0/c$, variability of the wind on a wide range
of time scales, $t_d<t_v<t_w$, is possible. For simplicity, we 
have assumed that in addition to $t_d$, 
which determines the initial shell thickness $\Delta=c t_d\sim R_0$, 
and $t_w$, the wind is characterized by a single
time scale $t_v>t_d$, which determines the shell ejection
rate $t_v^{-1}$.
We have addressed the questions of whether, and under what conditions, 
high radiative efficiency consistent with observations can be obtained, 
and whether the observed
clustering of peak emission energies can be naturally explained by the
model.

We have shown that a significant fraction, $\sim15\%$, of the fireball
energy can be converted to radiation. As pointed out at the end of \S2, 
our simplified treatment of post-collision shell evolution, assuming 
complete merger of shells, leads
to an overestimate of the reduction of the variance of colliding shells'
Lorenz factors with wind evolution. A more detailed calculation of the 
merger process will lead to an enhancement
of the radiative efficiency by a factor of order unity. The exact value
of the enhancement factor will depend, however, on the unknown internal
shell structure. 
Adopting simplifying assumptions regarding the 
shell structure and post-collision evolution, Kobayashi \& Sari (2001) have 
recently pointed out that including post-collsion evolution effects 
may lead to radiative efficiency exceeding $\epsilon_e$ 
(note, that this may be the case also when 
post-collision evolution is neglected). Although this effect may relax
somewhat the constraint imposed on $\epsilon_e$ by the requirement
of high efficiency, we have shown here
(see following paragraph), that $\epsilon_e$ is constrained to be of order
unity by the observed $\gamma$-ray spcetra.

In order to obtain high radiative efficiency and peak emission energy
$\sim1$~MeV, the minimum radius $R_i$ at which internal collisions in 
the expanding wind occur is required to be similar to $R_\pm$, 
the radius where the Thomson optical depth due to $e^\pm$ pairs
produced by shock synchrotron emission equals unity 
[see Eq. (\ref{rpmmin})], and a large variance (compared to the mean) of 
the colliding shells' Lorenz factor (LF) distribution 
is required. 
In addition, the electron and magnetic field energy fractions should
be close to equipartition, with $\epsilon_e\gtrsim0.1$ and
$\epsilon_B\gtrsim0.01$, 
in order for model predictions to be
consistent with the observed peak emission energy distribution 
[see Fig. 4 and Eq.(\ref{eq:emax1})].
The constraint $R_i\sim R_\pm$ is equivalent to a constraint on
the minimum LF, $\Gamma_m$, of expanding shells,   
$\Gamma_m\sim\Gamma_{m\pm}\approx10^{2.5}(L_{w,52}/t_{v,-3})^{2/9}$, where 
$L_w=10^{52}L_{w,52}{\rm erg\,s}^{-1}$ and $t_v=10^{-3}t_{v,-3}$~s
[see Eq. (\ref{eq:gpm})]. Large variance in the LF distribution of colliding
shells than requires a non-uniform LF distribution (e.g.
truncated log-normal or bimodal distributions, see Fig. 1)
with $\Gamma_M$, the maximum LF of wind shells,
close to the upper limit set by the shell acceleration  
process, $\Gamma<\eta_*\approx(\sigma_T L_w/4\pi m_p c^3 R_0)^{1/4}=
2\times10^3(L_{w,52}/R_{0,6})^{1/4}$, where
$R_0=10^6R_{0,6}$~cm.

Our results do not agree with the very high, $\sim80\%$,
radiative efficiency values, obtained in the calculations of Beloborodov
(2000) and Kobayashi \& Sari (2001) by assuming wind Lorentz factor
distributions extending from $\Gamma_m\sim10$ to $\Gamma_M\gtrsim10^4$. 
In our opinion the minimum wind Lorentz factor must be significantly
higher than 10 (as otherwise the wind becomes optically thick) while
the maximum Lorentz factor can not significantly exceed $10^3$ (due to
the acceleration process limitations).

We have shown that there is an upper limit to the observed energy 
of photons, $\varepsilon_p$,  at which the $\gamma$-ray flux peaks, 
$\varepsilon_p\lesssim\varepsilon_p^{\max}
\approx 0.5 R_{0,7}^{-5/6}$~MeV (where $R_0=10^7R_{0,7}$~cm), with very
weak dependence on $L_w$ and on $t_v$
[see discussion preceding Eq. (\ref{eq:emax1}), and figures 2, 5 and 6].
Thus, the source dynamical time $t_d\sim R_0/c$ must satisfy 
$t_d\lesssim 1$~ms in order to allow $\varepsilon_p\sim1$~MeV.
$\varepsilon_p\approx\varepsilon_p^{\max}$ is obtained for 
$\Gamma_m\approx\Gamma_{m\pm}$,
for which the radiative efficiency is largest, and 
$\varepsilon_p$ values in the range of 0.1~MeV to 1~MeV are obtained
for wind parameters for which the radiative efficiency is high, 
$\gtrsim10\%$ (see figure 3). 

High radiative efficiency and peak emission energy consistent with 
observations are therefore obtained for $\Gamma_m\sim\Gamma_{m\pm}$.
This does not necessarily imply that fine tuning of this model parameter
is required. For $\Gamma_m<\Gamma_{m\pm}$, most efficient
collisions occur at radii where the optical depth is high, leading
to low efficiency, and hence low luminosity, bursts with peak emission
energy $\sim1$~keV (see figures 2 and 3), which would not have been detected
by BATSE. For $\Gamma_m$ significantly
higher than $\Gamma_{m\pm}$, LF variance is small, leading to low
efficiency, low luminosity bursts with peak emission
energy $\sim10$~keV, which may have been difficult to detect with BATSE.
Natural consequences of the model considered here are therefore 
absence of bursts with peak emission energy significantly 
exceeding $\sim1$~MeV, and existence of low luminosity
bursts with low, $\sim1$~keV to $\sim10$~keV, peak emission energy.
The frequency of such bursts depends on the distribution
of $\Gamma_m$ in different winds.

It should be pointed out in this context that while the deficit, among
bursts detected by BATSE, of bursts
with peak emission energy below $\sim50$~keV clearly reflects a real deficit
of such bursts, the deficit in detected bursts with peak emission energy
$>0.5$~MeV may be partly due to a selection effect (e.g. \cite{Lloyd99}). 
BATSE data are consistent 
with equal number of bursts per logarithimic peak energy interval beyond
$\varepsilon_p\sim0.5$~MeV (with luminosity comparable to that
of lower peak energy bursts). Our model prediction, that bursts with
$\varepsilon_p\gg1$~MeV should
be absent, should therefore be tested with future GRB detectors, which
are able to better constrain the high energy end of the 
$\varepsilon_p$ distribution.

A note should be made here regarding the low energy spectral slope.
The upper limit $\eta_*$ on $\Gamma_M$ is a consequence of the fact that
for $\eta>\eta_*$ shell acceleration saturates at $\Gamma\sim\eta_*<\eta$,
as the wind becomes optically thin to Thomson scattering, and the internal
energy left in the shell escapes as thermal radiation rather than being 
converted to kinetic energy. The requirement $\Gamma\sim\eta_*$ therefore 
implies that a significant fraction of the wind energy may escape as thermal
radiation, leading to low energy spectral slopes steeper than those expected
for pure synchrotron emission. 
This may account, at list partially, for observed
steep low energy spectra. 

Finally we note that the lower limit imposed on $\Gamma_m$,
$\Gamma_m\gtrsim\Gamma_\pm$, is
not derived from the requirement that the pair production optical depth
for high energy, $>100$~MeV, photons be smaller than unity. While
this requirement leads to a similar constraint, $\Gamma_m\gtrsim10^2$,
high energy photons have been detected in a small number of cases only.
The constraint $\Gamma_m>\Gamma_{m\pm}$ is imposed in the present analysis 
by the requirement that the wind Thomson optical depth 
due to $e^\pm$ pairs produced by synchrotron photons be smaller
than unity at the internal shocks stage.

\acknowledgements{The research of DG and MS 
is supported by COFIN-99-02-02. EW is partially supported
by BSF Grant 9800343, AEC Grant 38/99 and MINERVA Grant.
We thank  Marco Salvati for useful comments on the manuscript.
D.G. and M.S. thank the Weizmann Institute of Science, where 
part of this research was carried out, for the hospitality and 
for the pleasant working atmosphere.}

\clearpage

\begin{figure*}
\vspace*{0.1cm}
\plotone{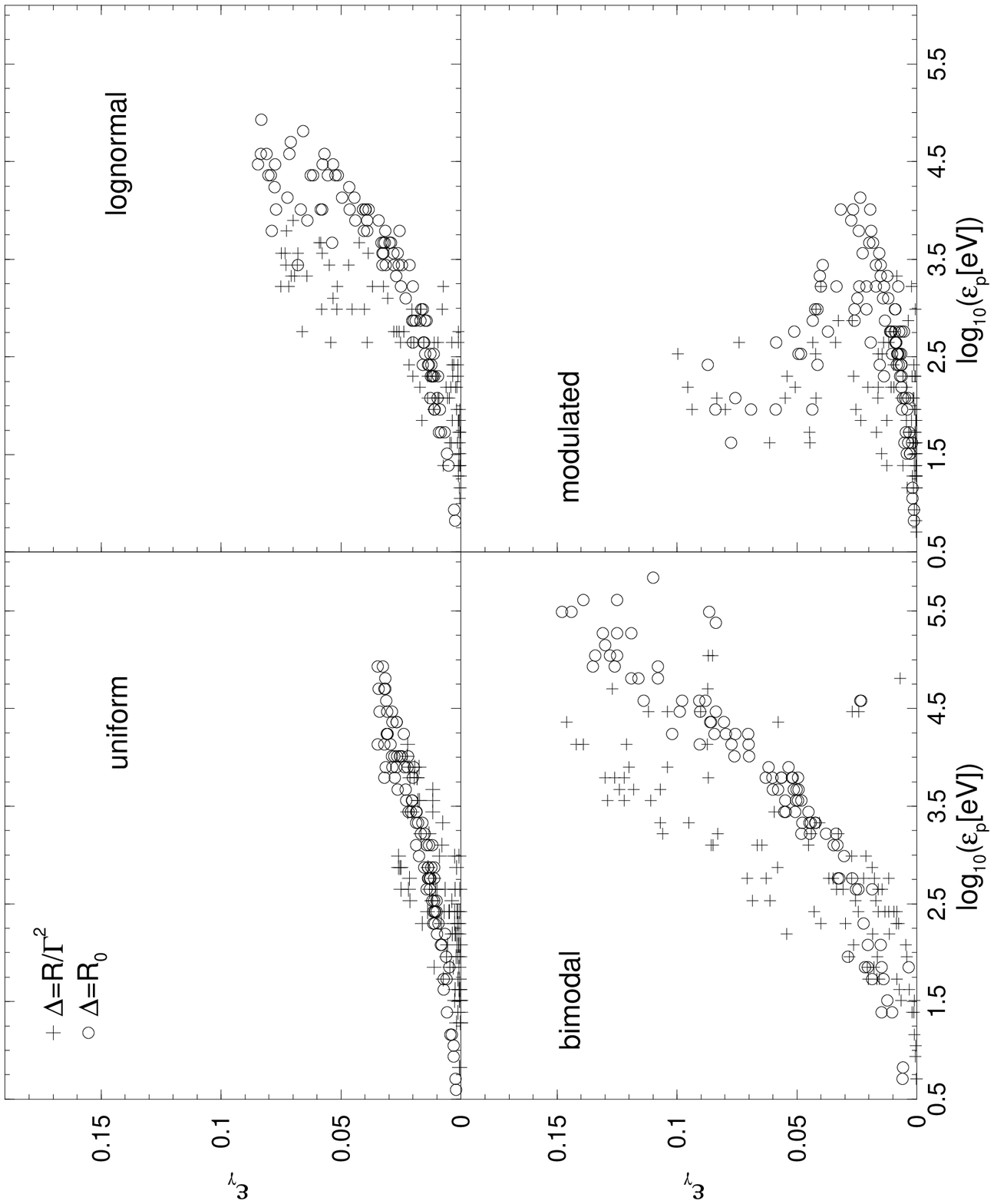}
\caption{Radiative efficiency versus peak emission energy
for various Lorenz factor distributions. 
Different points within each panel correspond to
different choices of $\Gamma_m$, $\Gamma_M$ and $t_v$. Results are shown
for $10< \Gamma_m < \Gamma_M < 2500$ and $10^{-4}{\rm\, s}<t_v<1$~s, 
for two extreme assumptions on shell width expansion,
maximal expansion $\Delta=\max(R_0,R/\Gamma^2)$ and constant width 
$\Delta=R_0$. We have assumed equal mass shells, 
$R_{0}=10^{6}$~cm, $L_{w}=10^{51}{\rm erg\, s}^{-1}$ and source redshift
$z=1$ for all calculations.  
 \label{fig:raggi}}
\end{figure*}

\begin{figure*}
\vspace*{0.1cm}
\plotone{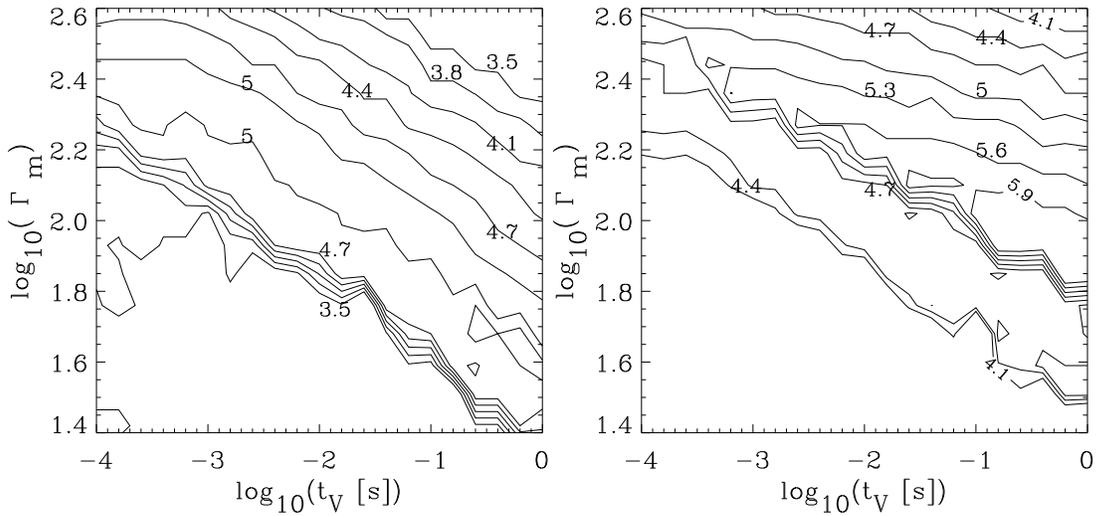}
\caption{Contour plots of the peak emission energy,
$\log_{10}(\varepsilon_p[{\rm eV}])$, as a function of
$t_v$ and the minimum Lorentz factor $\Gamma_m$
of the ejected shells. The value of $\Gamma_M$ is 2500 and $L_w=10^{52}$ erg
s$^{-1}$, the LF distribution is bimodal and the shells have equal masses.
The left panel refers to the case of expanding shell 
$\Delta=\max(R_0,R/\Gamma^2)$ and the right panel to the case
of time indipendent shell thickness $\Delta=R_0$.}
\end{figure*}

\begin{figure*}
\vspace*{0.1cm}
\plotone{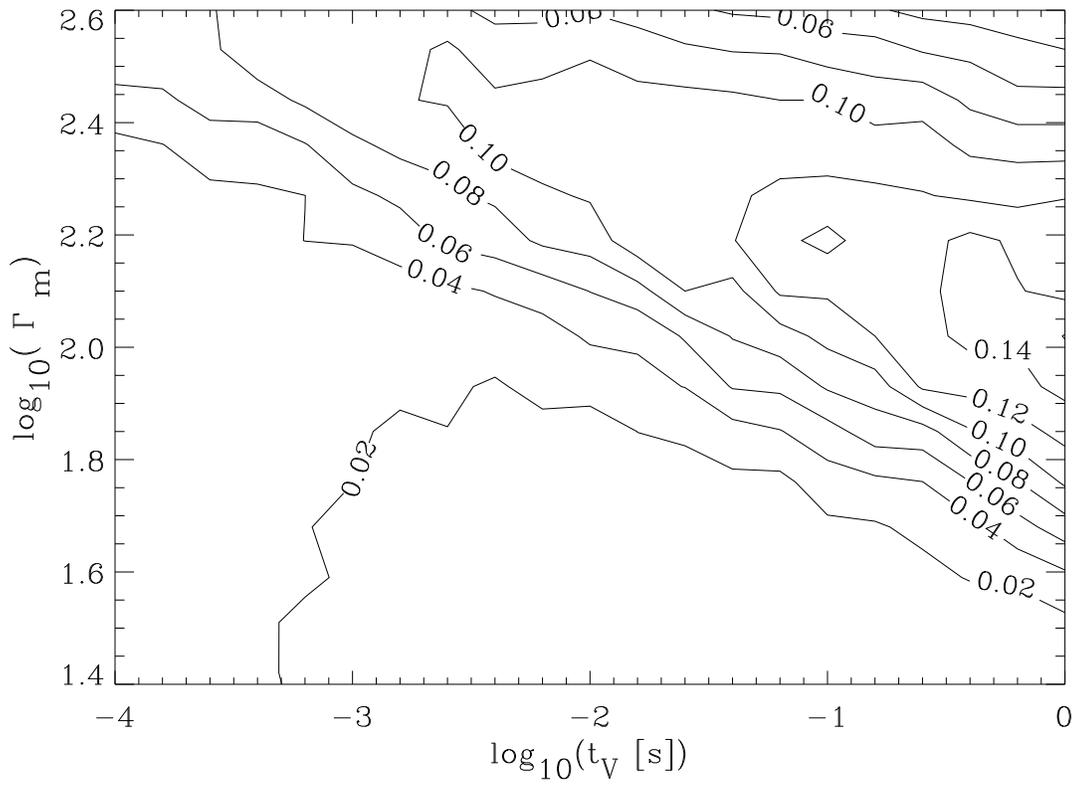}
\caption{Contour plot of the radiative efficiency for the case described in
Fig. 2. The radiative efficiency is similar for both
$\Delta=\max(R_0,R/\Gamma^2)$ and $\Delta=R_0$ cases.}
\end{figure*}

\begin{figure*}
\vspace*{0.1cm}
\plotone{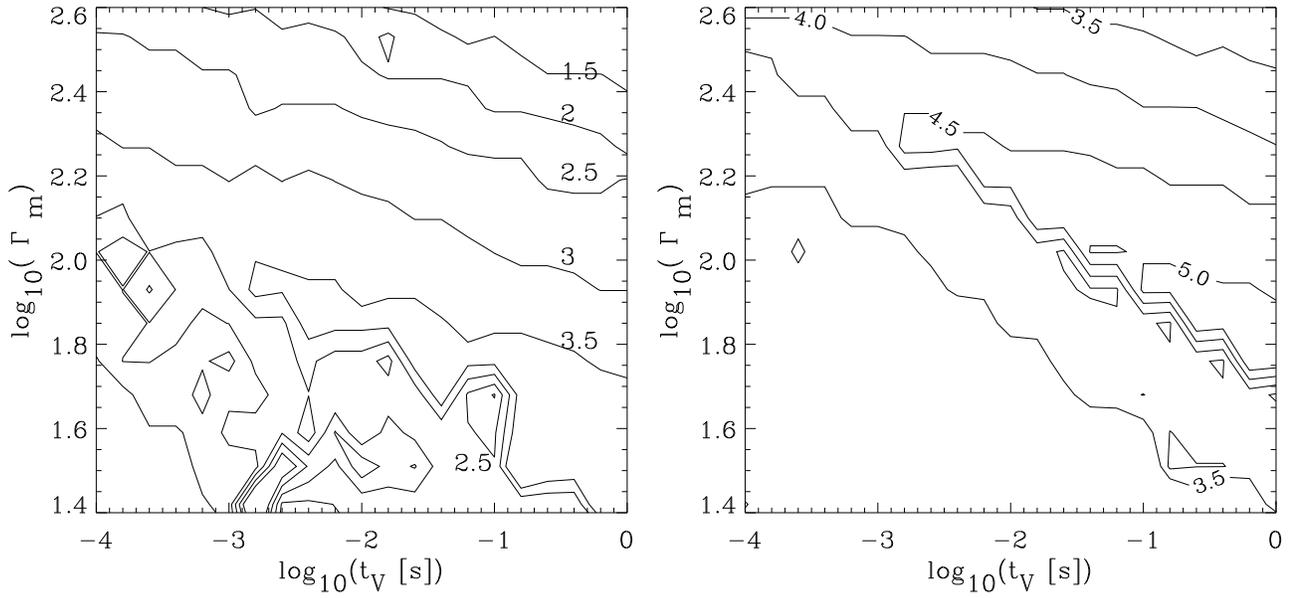}
\caption{Same as right panel of Fig. 2 (constant pre-shock shell width), 
for electron and magnetic field
energy fractions below equipartition. Left panel: $\epsilon_e=0.01$,
$\epsilon_B=0.1$; Right panel: $\epsilon_e=0.45$,
$\epsilon_B=0.001$.}
\end{figure*}

\begin{figure*}
\vspace*{0.1cm}
\plotone{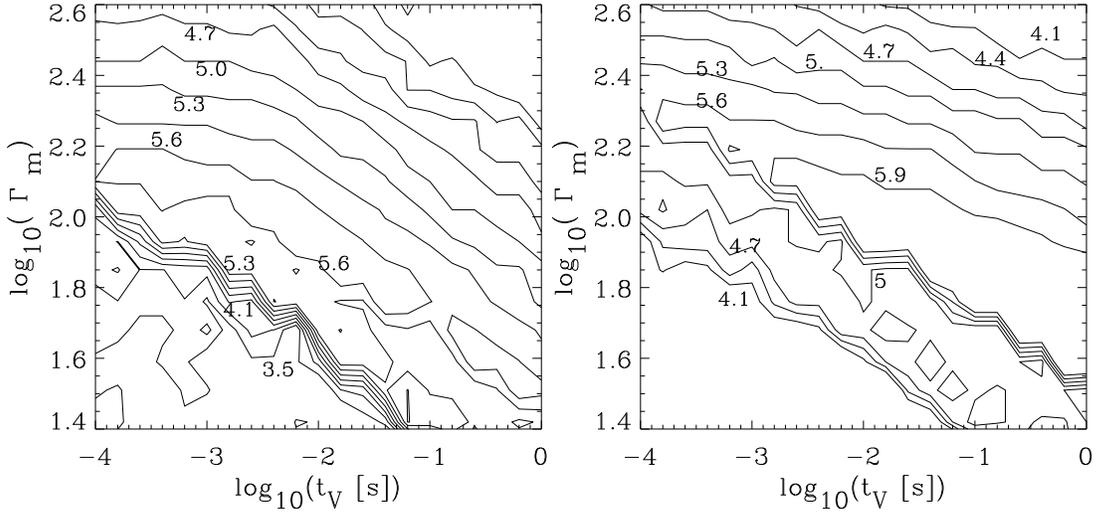}
\caption{Same as Fig. 2, for different luminosity, $L_w=10^{51}$ erg
s$^{-1}$.}
\end{figure*}

\begin{figure*}
\vspace*{0.1cm}
\plotone{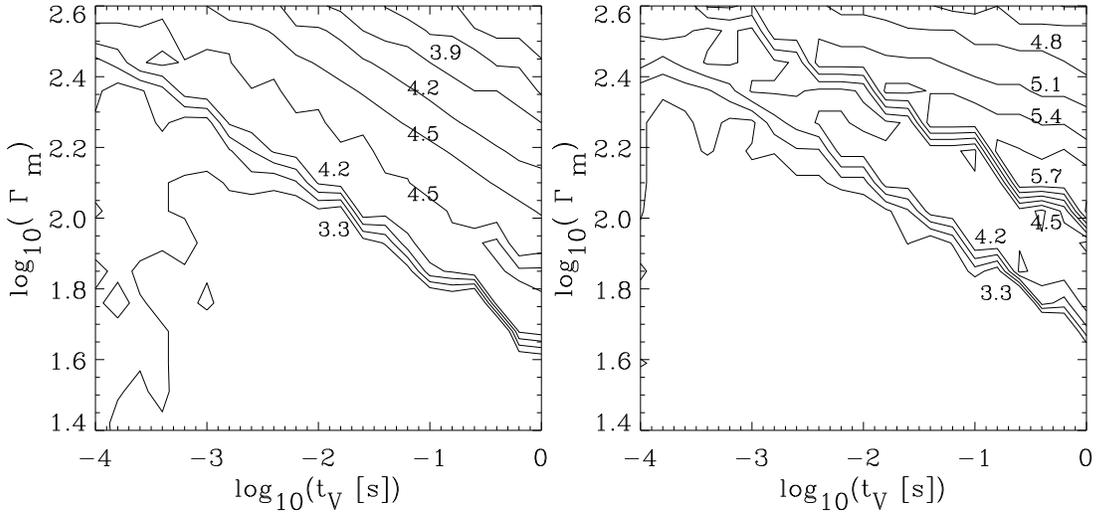}
\caption{ Same as Fig. 2, for $L_w=10^{53}$ erg s$^{-1}$.}
\end{figure*}
\clearpage

\end{document}